\documentstyle[12pt,epsfig]{article}

\begin{document}
\baselineskip=23pt

\vspace{1.2cm}

\begin{center}
{\Large \bf  Exact solutions of the Dirac equation in
Robertson-Walker space-time}

\bigskip

Xin-Bing Huang\footnote{huangxb@pku.edu.cn}\\
{\em Department of Physics,
Peking University,} \\
{\em    100871, Beijing, P.R.China}
\end{center}

\bigskip
\bigskip
\bigskip

\centerline{\large Abstract}

The covariant Dirac equation in Robertson-Walker space-time is
studied under the comoving coordinates. The exact forms of the
spatial factor of wave function are respectively acquired in
closed, spatially flat, and open universes.

 \vspace{1.2cm}


\vspace{1.2cm}

\newpage

\section{Introduction}
The Robertson-Walker metric plays a central role in modern
cosmology. The expanding universe is usually described by it.
Therefore the behavior of relativistic particles obeying the
covariant Dirac equation in Robertson-Walker space-time has evoked
many investigations in history. Such works go back to
Schr$\ddot{{\rm o}}$dinger~\cite{sch40}, Brill and
Wheeler~\cite{bw57}, Barut and Duru~\cite{bd87}. A general
discussion of the case of massless neutrino has been given by
Brill and Wheeler.

In literature, Li and Guo ever considered the exact solutions of
Klein-Gordon equation and the Dirac equation in Beltrami-de-Sitter
coordinates~\cite{lg82}. They gave right solutions for
Klein-Gordon equation and acquired the discrete
spectrum~\cite{ch02} of the mass of a spin-0 particle because of
the continuous condition of the Legend function at the boundary of
de Sitter space-time. But they obtained the wrong evolutionary
equation from the Dirac equation, thus gave the wrong solutions
for the Dirac equation in de Sitter space-time.

Had been aware of the importance of the behavior of a
spin-$\frac{1}{2}$ particle in the expanding universe, Barut and
Duru investigated the exact solutions of the Dirac equation for
three typical models of expanding universe~\cite{bd87}. But they
assumed that the space-time is spatially flat. Thus the exact
solution of the Dirac equation in general Robertson-Walker
space-time haven't been given in literature.

In this paper, we investigate the behavior of a massive
spin-$\frac{1}{2}$ particle in the open, spatially flat, and
closed Robertson-Walker space-time, respectively. The exact forms
of the spatial factor are obtained. Our work will be a correction
and extension to the paper of Li and Guo, on the other hand it is
also a complementary to the work of Barut and Duru. The exact
solutions of the Dirac equation in Robertson-Walker will give a
stage for studying the possible usage of neutrino
oscillations~\cite{xin04} in cosmology, we think.


\section{Dirac equation in Robertson-Walker space-time}

The cosmological principle, the hypothesis that all positions in
the universe are essentially equivalent, assumes that the
``smeared" universe is homogeneous and isotropic about every
point. The convenient implement to describe this ``smeared"
universe is the Robertson-Walker metric, which is manifestly
spherically symmetric. In the comoving coordinates, the
Robertson-Walker metric of the universe is taken to be\footnote{we
adopt the conventions that $x^{0}=t$, $x^{1}=r$, $x^{2}=\theta$,
$x^{3}=\varphi$ and $\hbar=c=1$.}
\begin{equation}
\label{RWmetric} ds^{2}={\rm
g}_{\mu\nu}dx^{\mu}dx^{\nu}=dt^{2}-R^{2}(t)\left[
\frac{dr^2}{1-kr^2}+r^2(d\theta^2+\sin^2\theta
d\varphi^2)\right]~,
\end{equation}
where $R(t)$ is an unknown function of time, which is called
cosmic scale factor, and $k$ is a constant, which by a suitable
choice of units for $r$ can be chosen to have the value $+1$, $0$,
or $-1$. When $k=+1,~0$, or $-1$, the space-time is respectively
closed, spatially flat or open.

To set the stage for solving the Dirac equation in a curved
space-time, we introduce the local tetrad field as
follows\footnote{In this paper,
 using Roman suffixes to
refer to the bases of local Minkowski frame; using Greek suffixes
to refer to curvilinear coordinates of space-time.}
\begin{equation}
\label{tetrad1} {\rm
g}^{\mu\nu}=e^{\mu}_{~a}(x)e^{\nu}_{~b}(x)\eta^{ab}~,
\end{equation}
where ${\rm g}^{\mu\nu}$ being the space-time metric, and the
metric tensor of Minkowski space-time $\eta_{ab}$ is
\begin{equation}
\label{metric102}
\eta^{00}=+1~,~~~~\eta^{11}=\eta^{22}=\eta^{33}=-1~,~~~~
\eta^{ab}=0~~~~{\rm for}~~~~a \neq b~.
\end{equation}

In a curved space-time, the covariant Dirac equation reads
\begin{equation}
\label{Diraceq} \left[ i\gamma^{\mu}(x)\frac{\partial}{\partial
x^{\mu}}- i\gamma^{\mu}(x)\Gamma_{\mu}(x)\right]\psi(x)=m\psi(x)~,
\end{equation}
Here $\gamma^{\mu}(x)$ are the curvature-dependent Dirac matrices
and $\Gamma_{\mu}(x)$ are the spin connections to be determined.
The curvature-dependent Dirac matrices $\gamma^{\mu}(x)$ are
presented in terms of the tetrad field as
\begin{equation}
\label{Diracmatrix} \gamma^{\mu}(x)=e^{\mu}_{~a}(x)\gamma^{a}~,
\end{equation}
where $\gamma^{a}$ denotes the standard flat-space Dirac matrices,
which satisfies $\{\gamma^{a},\gamma^{b}\}=2\eta^{ab}$.

The tetrad field in the case of Robertson-Walker metric
(\ref{RWmetric}) is thus taken to be
\begin{equation}
\label{tetrad2} e^{\mu}_{~a}(x)={\rm diag} \left(
1,\frac{\sqrt{1-kr^2}}{R(t)},\frac{1}{R(t)r},\frac{1}{R(t)r\sin\theta}
\right)~.
\end{equation}
Inserting Eq.(\ref{tetrad2}) into Eq.(\ref{Diracmatrix}) yields
\begin{eqnarray}\nonumber
\gamma^{0}(x)&=&\gamma_{0}~,~~~~~~~~~~~~~~\gamma^{1}(x)=-\frac{\sqrt{1-kr^2}}{R(t)}\gamma_{1}~,
\\
\label{con}
\gamma^{2}(x)&=&-\frac{1}{R(t)r}\gamma_{2}~,~~~~\gamma^{3}(x)=-\frac{1}{R(t)r\sin\theta}\gamma_{3}~.
\end{eqnarray}

The spin connections $\Gamma_{\mu}(x)$ in Eq.(\ref{Diraceq})
satisfy the equation
\begin{equation}
\label{spincon}
[\Gamma_{\mu}(x),\gamma^{\nu}(x)]=\frac{\partial\gamma^{\nu}(x)}{\partial
x^{\mu}}+\Gamma^{\nu}_{\mu\rho}\gamma^{\rho}(x)~,
\end{equation}
where $\Gamma^{\nu}_{\mu\rho}$ are the Christoffel symbols for the
metric (\ref{RWmetric}), which are determined by
\begin{equation}
\label{Christoffel} \Gamma^{\nu}_{\mu\rho}=\frac{1}{2}{\rm
g}^{\nu\tau} \left(\frac{\partial {\rm g}_{\tau\rho}}{\partial
x^{\mu}}+\frac{\partial {\rm g}_{\mu\tau}}{\partial
x^{\rho}}-\frac{\partial {\rm g}_{\mu\rho}}{\partial x^{\tau}}
\right)~.
\end{equation}
Hence Eq.(\ref{spincon}) can be solved for the spin connections
$\Gamma_{\mu}(x)$ which we determine as
\begin{eqnarray}\nonumber
\Gamma_{0}&=&0~,~~~~~~~\Gamma_{1}=\frac{\dot{R}(t)}{2\sqrt{1-kr^2}}\gamma_{0}\gamma_{1}~,
\\
\label{spinconnection}
\Gamma_{2}&=&\frac{1}{2}\dot{R}(t)r\gamma_{0}\gamma_{2}+\frac{1}{2}\sqrt{1-kr^2}\gamma_{2}\gamma_{1}~,
\\
\nonumber
\Gamma_{3}&=&\frac{1}{2}\dot{R}(t)r\sin\theta\gamma_{0}\gamma_{3}
+\frac{1}{2}\sqrt{1-kr^2}\sin\theta\gamma_{3}\gamma_{1}+\frac{1}{2}\cos\theta\gamma_{3}\gamma_{2}~,
\end{eqnarray}
where $\dot{R}(t)$ denotes $\frac{d R(t)}{dt}$, so that the
combination $\gamma^{\mu}(x)\Gamma_{\mu}(x)$ in Eq.(\ref{Diraceq})
simplifies to
\begin{equation}
\label{spinconsum}
\gamma^{\mu}(x)\Gamma_{\mu}(x)=-\frac{3\dot{R}}{2R}\gamma_{0}+\frac{\sqrt{1-kr^2}}{R(t)r}\gamma_{1}
+\frac{\cos\theta}{2R(t)r\sin\theta}\gamma_{2}~.
\end{equation}

Inserting Eq.(\ref{con}) and Eq.(\ref{spinconsum}) into
Eq.(\ref{Diraceq}), then we obtain the Dirac equation in
Robertson-Walker space-time
\begin{equation}\label{diraceq2}
\begin{array}{l}
\displaystyle\left\{ iR\gamma_{0} \left( \frac{\partial}{\partial
t}+\frac{3\dot{R}}{2R} \right) - i \left[ \sqrt{1-kr^2}\gamma_{1}
\left(\frac{\partial}{\partial r}+\frac{1}{r} \right)
+\frac{1}{r}\gamma_{2}\left(\frac{\partial}{\partial\theta}+
\frac{\cos\theta}{2\sin\theta}
\right)\right.\right.\\[1cm]
\displaystyle~~~~~~~~~~~\left.\left.+\frac{1}{r\sin\theta}\gamma_{3}\frac{\partial}{\partial\varphi}
\right]- m R \right\} \psi(t,r,\theta,\varphi)=0~.
\end{array}
\end{equation}


\section{The exact solutions of the Dirac equation}
We can simplify the Dirac equation (\ref{diraceq2}) by
writing~\cite{lg82}
\begin{equation}
\label{seperation}
\psi(t,r,\theta,\varphi)=(\sin\theta)^{-\frac{1}{2}}
R^{-\frac{3}{2}}\Psi(t,r,\theta,\varphi)~.
\end{equation}
The reduced equation in terms of $\Psi(t,r,\theta,\varphi)$ is of
the form
\begin{equation}
\label{neweq} R\left( \frac{\partial}{\partial t }+ i m \gamma_{0}
\right)\Psi=\left[\sqrt{1-kr^2}\gamma_{0}\gamma_{1}\left(\frac{\partial}{\partial
r}+\frac{1}{ r}\right)+\frac{1}{
r}\gamma_{1}\hat{K}(\theta,\varphi)\right]\Psi~,
\end{equation}
where
\begin{equation}
\label{Kdef} \hat{K}(\theta,\varphi)=\gamma_{0}\gamma_{1}\left(
\gamma_{2}\frac{\partial}{\partial\theta}+\gamma_{3}\frac{1}{\sin\theta}\frac{\partial}{\partial\varphi}
\right)
\end{equation}
is a Hermitian operator, as we know, which is defined in
Ref.\cite{sch38} first. The eigen equation of $\hat{K}$ is
\begin{equation}
\label{Keigen} \hat{K}\Psi_{\varsigma,\varepsilon}=
\varsigma\Psi_{\varsigma,\varepsilon}~,~~~~~~~~\varsigma=0, \pm 1,
\pm 2, \cdot\cdot\cdot,
\end{equation}
where the eigen value $\varsigma$ is an integer, which has been
proven in Ref.\cite{bw57,sch38}.

Only the two matrices $\gamma_{0}$ and $\gamma_{1}$ remain
explicitly in the simplified Dirac equation (\ref{neweq}), they
can therefore be represented by $2\times2$ matrices
\begin{equation}
\label{twocomponent} \gamma_{0}=\left(
\begin{array}{cc}
{\bf 1} & 0
\\
0& -{\bf 1}
\end{array}\right)~,~~~\gamma_{1}=\left(
\begin{array}{cc}
0 & -\tau_{1}
\\
\tau_{1} & 0
\end{array}\right)~,
\end{equation}
where ${\bf 1}$, $\tau_{1}$ are respectively $2\times 2$ unit
matrix and the Pauli matrix, namely
\begin{equation}
\label{paulima} \tau_{1}=\left(
\begin{array}{cc}
0 & 1
\\
1& 0
\end{array}\right)~.
\end{equation}
We separate the angular factor from the wave function, and
represent the radial and temporal factor by a two-component
spinor, that is
\begin{equation}\label{sep}
\Psi_{\varsigma,\varepsilon}=\Theta_{\varsigma}(\theta,\varphi)
\left(
\begin{array}{c}
\phi(r,t)
\\
\chi(r,t)
\end{array}
\right)~,
\end{equation}
where the angular factor, $\Theta_{\varsigma}(\theta,\varphi)$, is
determined by the requirement
$\hat{K}\Theta_{\varsigma}(\theta,\varphi)=\varsigma
\Theta_{\varsigma}(\theta,\varphi)$. This equation for eigenstates
of the angular motion has been investigated by Schr$\ddot{\rm
o}$dinger~\cite{sch38}. He found, the operator $\hat{K}$ is
related to the total angular momentum. Their eigenvalues
satisfy~\cite{gre00}
\begin{equation}
\label{egrelation} \displaystyle
\varsigma=\mp(j+\frac{1}{2})=\left\{
\begin{array}{l}
\displaystyle -(l+1)~~~~{\rm for}~~j=l+\frac{1}{2}
\\[0.2 cm]
\displaystyle l~~~~~~~~~~~~~~{\rm for}~~j=l-\frac{1}{2}
\end{array}
\right.~.
\end{equation}
Also the eigenfunction $\Theta_{\varsigma}$ is related with the
spherical harmonics as follows
\begin{equation}
\label{har} \Theta_{\varsigma}(\theta,\varphi)\propto Y_{l,{\bar
m}+\frac{1}{2}}(\theta,\varphi)\propto P^{{\bar
m}+\frac{1}{2}}_{l}(\cos\theta)e^{i({\bar
m}+\frac{1}{2})\varphi}~,
\end{equation}
or
\begin{equation}
\label{har2} \Theta_{\varsigma}(\theta,\varphi)\propto Y_{l,{\bar
m}-\frac{1}{2}}(\theta,\varphi)\propto P^{{\bar
m}-\frac{1}{2}}_{l}(\cos\theta)e^{i({\bar
m}-\frac{1}{2})\varphi}~,
\end{equation}
where $P(\cos\theta)$ being the associated Legendre functions, and
${\bar m}\pm\frac{1}{2}=0,\pm 1,\pm 2,\cdot\cdot\cdot,\pm l$.

Using equations (\ref{Keigen}), (\ref{twocomponent}), and
(\ref{sep}), we can rewrite the equation (\ref{neweq}) into two
equations
\begin{eqnarray}
\label{dir1} && R\left( \frac{\partial}{\partial t }+ i m
 \right)\phi(r,t)=\left[\sqrt{1-kr^2}\left(\frac{\partial}{\partial
r}+\frac{1}{r}\right)+\frac{\varsigma}{
r}\right]\tau_{1}\chi(r,t)~,
\\
\label{dir2} && R\left( \frac{\partial}{\partial t }- i m
 \right)\chi(r,t)=\left[\sqrt{1-kr^2}\left(\frac{\partial}{\partial
r}+\frac{1}{r}\right)-\frac{\varsigma}{
r}\right]\tau_{1}\phi(r,t)~.
\end{eqnarray}
To solve the above equations, we separate the functions
$\phi(t,r)$ and $\chi(t,r)$ into radial and temporal factors,
respectively
\begin{equation}
\label{septr}
\phi(t,r)=U_{1}(r)T_{1}(t)~,~~~~\chi(t,r)=U_{2}(r)T_{2}(t).
\end{equation}
With the help of Eq.(\ref{paulima}), inserting the above equations
into Eq.(\ref{dir1}) and Eq.(\ref{dir2}) yields the evolution
equations
\begin{eqnarray}
\label{eveqphi} &&  R\left( \frac{d}{d t }+ i m
 \right)T_{1}(t)-i\varepsilon T_{2}(t)
=0~,
\\
\label{eveqchi} && R\left( \frac{d}{d t }- i m
 \right)T_{2}(t)-i\varepsilon T_{1}(t)
=0~,
\end{eqnarray}
and the radial equations
\begin{eqnarray}
\label{spatialeqphi} && \left[\sqrt{1-kr^2}\left(\frac{d}{d
r}+\frac{1}{ r}\right)+\frac{\varsigma}{
r}\right]\left[\sqrt{1-kr^2}\left(\frac{d}{d
r}+\frac{1}{r}\right)-\frac{\varsigma}{
r}\right]U_{1}+\varepsilon^{2}U_{1}=0~,
\\
\label{spatialeqchi} && \left[\sqrt{1-kr^2}\left(\frac{d}{d
r}+\frac{1}{ r}\right)-\frac{\varsigma}{
r}\right]\left[\sqrt{1-kr^2}\left(\frac{d}{d
r}+\frac{1}{r}\right)+\frac{\varsigma}{
r}\right]U_{2}+\varepsilon^{2}U_{2}=0~.
\end{eqnarray}
Obviously, if $U_{1}$ is substituted by $U_{2}$, and
simultaneously $\varsigma$ are replaced by $-\varsigma$, then the
equation (\ref{spatialeqphi}) becomes the equation
(\ref{spatialeqchi}). Therefore, in the following, it is of only
necessity to study the equation (\ref{spatialeqphi}).

The equation (\ref{spatialeqphi}) directly simplifies to
\begin{equation}
\label{simple} (1-kr^2)\frac{d^2 U_{1}}{d
r^2}+\left(\frac{2}{r}-3k r \right)\frac{d U_{1}}{d
r}+\left(\frac{\varsigma\sqrt{1-kr^2}-\varsigma^2}{
r^2}+\varepsilon^{2}-k \right)U_{1}=0~.
\end{equation}
The constant $k$ appears in the above equation, the value of which
will affect the exact solution of Eq.(\ref{simple}). We,
therefore, investigate Eq.(\ref{simple}) according to the value of
$k$ respectively.

\subsection{In spatially flat Robertson-Walker geometry}
In the case of spatially flat Robertson-Walker geometry, the
constant $k$ satisfies $k=0$. The equation (\ref{simple}) is thus
reduced to
\begin{equation}
\label{simplek0} \frac{d^2 U_{1}}{d r^2}+\frac{2}{r}\frac{d
U_{1}}{d r}+\left(\frac{\varsigma(1-\varsigma)}{
r^2}+\varepsilon^{2} \right)U=0~.
\end{equation}
Through a simple transformation, one can easily find that the
equation (\ref{simplek0}) is the Bessel's equation of order
$-\varsigma+\frac{1}{2}$. The solutions of the above equation
are~\cite{wg00}
\begin{equation}
\label{solution00} U_{1}(r)=\frac{1}{\sqrt{\varepsilon
r}}J_{\pm(\varsigma-\frac{1}{2})}(\varepsilon r)~.
\end{equation}
Where $J_{\pm(\varsigma-\frac{1}{2})}(\varepsilon r)$ are called
the Bessel functions.

\subsection{In closed Robertson-Walker geometry}
In the case of $k=+1$, the Robertson-Walker space-time is closed.
The equation (\ref{simple}) becomes
\begin{equation}
\label{simplek+1} (1-r^2)\frac{d^2 U_{1}}{d
r^2}+\left(\frac{2}{r}-3 r \right)\frac{d U_{1}}{d
r}+\left(\frac{\varsigma\sqrt{1-r^2}-\varsigma^2}{
r^2}+\varepsilon^{2}-1 \right)U_{1}=0~.
\end{equation}
Do the transformation
\begin{equation}
\label{trans+1} \rho=\frac{\sqrt{1-r^2}-1}{\sqrt{1-r^2}+1}~,
\end{equation}
then, in terms of $\rho$, the radial equation (\ref{simplek+1}) is
rewritten as follows
\begin{equation}
\label{sim+1} \rho(1-\rho)^2\frac{d^2 U_{1}}{d
\rho^2}+\frac{1}{2}(1-\rho)(3-\rho)\frac{d U_{1}}{d
\rho}+\left[\frac{\varsigma(1-\rho)}{2\rho}-
\frac{(\varsigma+\varsigma^2)(1-\rho)^2}{4\rho}+1-\varepsilon^{2}
\right]U_{1}=0~.
\end{equation}
The above equation can be transformed into a hypergeometric
equation, In the interval of $0\le r<1$, the solutions of which
can thus be represented as~\cite{wg00}
\begin{eqnarray}
\nonumber
U_{1}(r,\varsigma>0)&=&\left(\frac{\sqrt{1-r^2}-1}{\sqrt{1-r^2}+1}
\right)^{\frac{\varsigma-1}{2}}\left( \frac{2}{\sqrt{1-r^2}+1}
\right)^{1+\varepsilon}\\
\nonumber &&\cdot
F\left(\varsigma+\varepsilon+\frac{1}{2},\varepsilon,\varsigma+\frac{1}{2};
\frac{\sqrt{1-r^2}-1}{\sqrt{1-r^2}+1} \right)~,
\\
\nonumber U_{1}(r,\varsigma \leq
0)&=&\left(\frac{\sqrt{1-r^2}-1}{\sqrt{1-r^2}+1}
\right)^{-\frac{\varsigma}{2}}\left( \frac{2}{\sqrt{1-r^2}+1}
\right)^{1+\varepsilon}\\
\label{solution+1} &&\cdot
F\left(1+\varepsilon,\frac{1}{2}+\varepsilon-\varsigma,\frac{3}{2}-\varsigma;
\frac{\sqrt{1-r^2}-1}{\sqrt{1-r^2}+1} \right)~.
\end{eqnarray}
Where $F$ is the hypergeometric function.

\subsection{In open Robertson-Walker geometry}
In this subsection, we investigate the case of $k=-1$, where the
space-time is open. The radial equation (\ref{simple}) is
simplified to
\begin{equation} \label{simplek-1}
(1+r^2)\frac{d^2 U_{1}}{d r^2}+\left(\frac{2}{r}+3 r
\right)\frac{d U_{1}}{d
r}+\left(\frac{\varsigma\sqrt{1+r^2}-\varsigma^2}{
r^2}+\varepsilon^{2}+1 \right)U_{1}=0~.
\end{equation}
Do the similar transformation as Eq.(\ref{trans+1}), set
\begin{equation}
\label{trans-1} \rho=\frac{\sqrt{1+r^2}-1}{\sqrt{1+r^2}+1}~.
\end{equation}
We then rewrite the equation (\ref{simplek-1}) as
\begin{equation}
\label{sim-1} \rho(1-\rho)^2\frac{d^2 U_{1}}{d
\rho^2}+\frac{1}{2}(1-\rho)(3-\rho)\frac{d U_{1}}{d
\rho}+\left[\frac{\varsigma(1-\rho)}{2\rho}-
\frac{(\varsigma+\varsigma^2)(1-\rho)^2}{4\rho}+\varepsilon^{2}+1
\right]U_{1}=0~.
\end{equation}
Similar to the equation (\ref{sim+1}), the above equation can also
be transformed into a hypergeometric equation~\cite{wg00}, the
solutions of which are taken to be
\begin{eqnarray}
\nonumber
U_{1}(r,\varsigma>0)&=&\left(\frac{\sqrt{1+r^2}-1}{\sqrt{1+r^2}+1}
\right)^{\frac{\varsigma-1}{2}}\left( \frac{2}{\sqrt{1+r^2}+1}
\right)^{1+i\varepsilon}\\
\nonumber &&\cdot
F\left(\varsigma+i\varepsilon+\frac{1}{2},i\varepsilon,\varsigma+\frac{1}{2};
\frac{\sqrt{1+r^2}-1}{\sqrt{1+r^2}+1} \right)~,
\\
\nonumber U_{1}(r,\varsigma \leq
0)&=&\left(\frac{\sqrt{1+r^2}-1}{\sqrt{1+r^2}+1}
\right)^{-\frac{\varsigma}{2}}\left( \frac{2}{\sqrt{1+r^2}+1}
\right)^{1+i\varepsilon}\\
\label{solution-1} &&\cdot
F\left(1+i\varepsilon,\frac{1}{2}+i\varepsilon-\varsigma,\frac{3}{2}-\varsigma;
\frac{\sqrt{1+r^2}-1}{\sqrt{1+r^2}+1} \right)~.
\end{eqnarray}

We have indicated that the solution of $U_{2}$ can be obtained
from equations (\ref{solution00}), (\ref{solution+1}), and
(\ref{solution-1}) by substituting $-\varsigma$ for $\varsigma$.
Hence we have acquired the exact forms of the spatial factor of
wave function in Robertson-Walker geometry.

In a different coordinates of spatially flat Robertson-Walker
space-time, the authors of Ref.\cite{bd87} have investigated the
exact solutions of the Dirac equation for three models of
expanding universes. It is obvious that our evolutionary equations
(\ref{eveqphi}) and (\ref{eveqchi}) are equivalent to the equation
(10) in Ref.\cite{bd87}. Hence the solutions of equations
(\ref{eveqphi}) and (\ref{eveqchi}) must be the same as the
results given in Ref.\cite{bd87}.

The authors of Ref.\cite{bd87} have given the exact solutions of
three special models for expansion: 1. $R=a_{0}t$, $a_{0}$ being a
constant, a model of a linear expanding universe; 2.
$R=a_{0}\sqrt{t}$, a model of a radiation-dominated universe; 3.
$R=e^{Ht}$, $H$ being a constant also, a model of de Sitter
universe. These models are typical in cosmology, we will not
discuss any more.

\section{Conclusion}

The behavior of the spin-$\frac{1}{2}$ particles in
Robertson-Walker space-time is investigated. The exact expressions
for the spatial factor of the spin-$\frac{1}{2}$ wave functions in
closed, spatially flat, and open Robertson-Walker space-time are
given, respectively. This work is a complementary to the
Ref.\cite{bd87}.

{\bf Acknowledgement:}  I would like to thank Prof. D. Du, Prof.
C. J. Zhu, Prof. Z. Chang, Prof. A. Sugamoto and Prof. I. Oda for
their help and encouragement.

\end{document}